\documentclass[runningheads]{llncs}
\usepackage{graphicx}
\usepackage{subfigure}
\usepackage{amsmath}
\usepackage{amssymb}
\usepackage{wrapfig}
\usepackage{upgreek}

\newcommand{\PP}{\mathbb{P}}
\newcommand{\GG}{\mathbb{G}_1}
\newcommand{\FF}{\mathbb{G}_2}
\begin{document}
\title{Going Beyond Saliency Maps:\\Training Deep Models to Interpret Deep Models}
\titlerunning{Deep Visualization}
%
\author{Zixuan Liu\inst{1} \and
Ehsan Adeli\inst{2,3} \and
Kilian M. Pohl\inst{2,4} \and
Qingyu Zhao\inst{2}}
\authorrunning{Liu et al.}
\institute{Department of Electrical Engineering, Stanford University, CA 94305 \and Department of Psychiatry \& Behavioral Sciences, Stanford University, CA 94305 \and Department of Computer Science, Stanford University, CA 94305 \and Center for Biomedical Sciences, SRI International, CA 94025}

\maketitle              
\begin{abstract} 
Interpretability is a critical factor in applying complex deep learning models to advance the understanding of brain disorders in neuroimaging studies. To interpret the decision process of a trained classifier, existing techniques typically rely on \textit{saliency maps} to quantify the voxel-wise or feature-level importance for classification through partial derivatives. Despite providing some level of localization, these maps are not human-understandable from the neuroscience perspective as they often do not inform the specific type of morphological changes linked to the brain disorder. Inspired by the image-to-image translation scheme, we propose to train simulator networks to inject (or remove) patterns of the disease into a given MRI based on a warping operation, such that the classifier increases (or decreases) its confidence in labeling the simulated MRI as diseased. To increase the robustness of training, we propose to couple the two simulators into a unified model based on \textit{conditional convolution}. We applied our approach to interpreting classifiers trained on a synthetic dataset and two neuroimaging datasets to visualize the effect of Alzheimer's disease and alcohol dependence. Compared to the saliency maps generated by baseline approaches, our simulations and visualizations based on the Jacobian determinants of the warping field reveal meaningful and understandable patterns related to the diseases.

\end{abstract}
\section{Introduction}
In recent years, deep learning has achieved unparalleled success in the field of medical image computing \cite{zhu2019} and is increasingly used to classify patients with brain diseases from normal controls based on their Magnetic Resonance Imaging (MRI) data \cite{Lee2019}. Compared to traditional machine learning methods, deep models can generally result in superior classification accuracy  \cite{Willemink2020} by training on a large amount of raw imaging data and employing more complex network architectures and learning strategies.
However, a primary challenge of applying complex deep networks to 3D MRI data is the lack of \textit{model interpretability}, which arguably plays a more pivotal role compared to the prediction accuracy itself. For example, when a learning model is used to aid the diagnosis by human experts, one needs to understand how the model reasons its prediction \cite{Brammer2009}. In other studies where neuroimaging is not a part of the diagnosis workflow (i.e., discovery-oriented analysis), the goal of learning image-based classifiers is solely for revealing the impact of the disease on the brain \cite{ouyang2020}. 

Compared to the large body of literature on model development, methods for model interpretation (or model visualization) are either oversimplified or misspecified for neuroimaging studies. For example, the most widely used visualization techniques to date are gradient-based methods \cite{springenberg2014striving,selvaraju2017grad}, which aim to generate a \textit{saliency map} for a given MRI. This map encodes the importance of the information contained within each voxel (or local neighborhood) in driving the model prediction. 
Despite the wide usage in computer vision tasks, the application of gradient-based methods in neuroimaging studies is limited as the saliency maps are generally noisy on the voxel level, imprecise in locating object boundaries, and applicable to only selective network architectures. Most importantly, the saliency maps only indicate the location of brain structures impacted by the disease but do not inform what type of morphological changes are induced (e.g., atrophy of cortical gray matter associated with Alzheimer's disease). 

In this work, we propose to interpret a trained classifier by learning two additional simulator networks, which aim to learn human-understandable morphological patterns within an image that can impact the classifier’s prediction. Motivated by the image-to-image translation scheme, one simulator warps an MRI of a healthy subject to inject the disease pattern such that the classifier increases its confidence in predicting the MRI as diseased (i.e., logit shift), and the other simulator removes the patterns from the MRI of a diseased subject to decrease the confidence. We then visualize the disease pattern on a subject-level by comparing the image appearance between the raw and simulated image pair or by quantifying the Jacobian map (encoding tissue expansion and shrinkage) of the warping field. To generate robust simulators, we employ a cycle-consistent scheme to encourage the simulators to inject and remove patterns only related to disease while preserving subject-specific information irrelevant to the disease. Furthermore, we propose to couple the two simulators into one coherent model by using the conditional convolution operation. The proposed visualization method was applied to interpret classifiers trained on a synthetic dataset, 1344 T1-weighted MRIs from the Alzheimer's Disease Neuroimaging Initiative (ADNI), and a dataset of 1225 MRIs for classifying individuals with alcohol dependence. We compared our visualization with a number of widely used alternative  techniques. Unlike the visualizations from those alternatives, our learning-based method generated images that capture high-level shape changes of brain tissue associated with the disease.

\section{Related Work}
Most existing methods for model interpretation visualize feature-level or voxel-level saliency scores quantified by partial derivatives \cite{springenberg2014striving,selvaraju2017grad}; i.e., how a small change in feature or voxel values can influence the final prediction. These derivatives can be efficiently computed by back-propagation. However, the voxelwise derivatives are generally noisy and non-informative as the variation in a voxel is more related to low-level features rather than the final prediction. One of the exceptions is Grad-CAM \cite{selvaraju2017grad}, which can generate smooth and robust visualization based on deriving feature-level importance but cannot accurately locate object boundaries and is only applicable to certain types of networks. Other than using partial derivatives, occlusion-based methods \cite{Zeiler2013} quantify the importance of a local image neighborhood for prediction by first masking the regional information in the images  (zero-out, blur, shuffle, deformation, etc.) and then evaluating the impact on the classifier’s prediction accuracy. However, the resulting saliency map can only be defined for the whole population but not for each individual. Recently, Ghorbani \cite{ghorbani2019automatic} has proposed a concept-based interpretation, which aims to directly identify critical image segments that drive the model decision. However, when applied to neuroimaging applications, all the above methods can only locate brain structures that are important for prediction but do not explain the alteration of those structures associated with a disease.

Recently, image-to-image translation frameworks have achieved marked success in medical applications including denoising, multi-modal image registration, and super-resolution reconstruction \cite{kaji2019overview,qin2019unsupervised}. The goal of such frameworks is to learn a bijective mapping between two distributions from different image domains. Inspired by this technique, we formulate the two domains as MRIs of healthy and diseased cohorts and learn how an MRI of a healthy subject will be altered if the subject is affected by the disease (and vice versa).

\section{Methods}
\begin{figure}[!t]
\centering
\includegraphics[width=\textwidth,trim=10 250 150 95,clip]{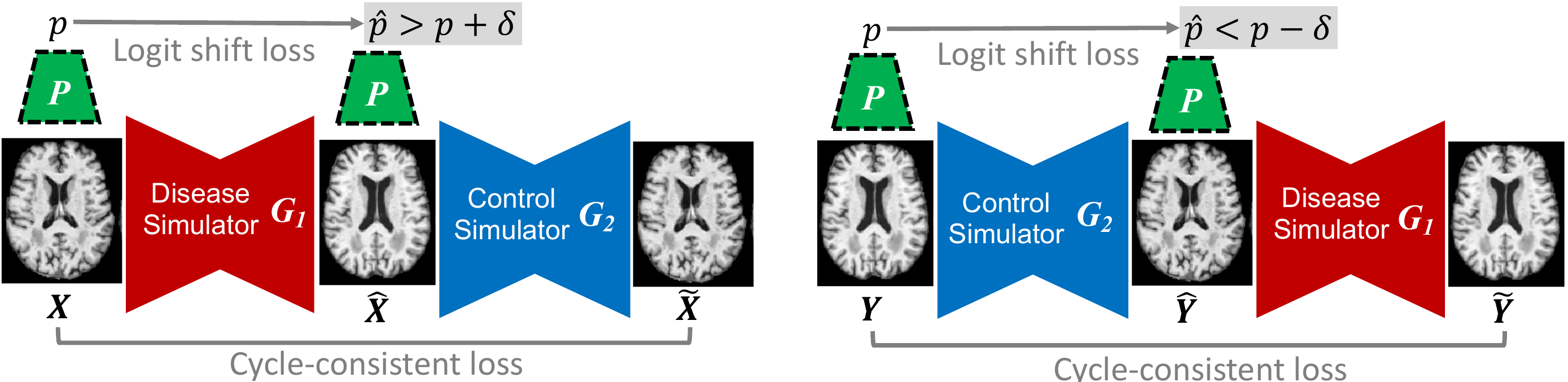}
\caption{To interpret a trained classifier ($\PP$), the proposed framework trains two separate simulator networks to learn the morphological change that defines the separation between control and diseased cohorts. The disease simulator ($\GG$) injects the disease pattern into an image $X$ such that the prediction logit $\hat{p}$ for the simulated image $\hat{X}$ increases by a pre-dined threshold $\delta$. The control simulator removes the pattern from an image. The two simulators are trained in a cycle-consistent fashion.} 
\label{fig:overview}
\end{figure}
\subsection{Cycle-Consistent Image Simulation}
Let $\boldsymbol{\mathcal{X}}$  be the set of MRI of controls and $\boldsymbol{\mathcal{Y}}$ the set of diseased participants. We assume a deep classifier $\PP$ has been trained on the datasets such that $p = \PP(X)$ is the \textit{logit} (value before \textit{sigmoid}) encoding the confidence in labeling image $X$ as diseased; i.e., $X$ is from a diseased subject if $p>0$ and from a control subject otherwise. Now our goal is to visualize the model $\PP$ to understand how the morphological information in $X$ impacts the final prediction $p$. To do so, we propose to train two simulator networks $\GG$ and $\FF$, where $\GG(\cdot)$ alters the MRI of a control subject $X \in \boldsymbol{\mathcal{X}}$ to resemble a diseased one $\hat{X} := \GG(X)$ (adding the disease pattern to $X$ as if the subject was affected by the disease), and $\FF(\cdot)$ removes the disease pattern from an MRI $Y \in \boldsymbol{\mathcal{Y}}$ so that $\hat{Y} := \FF(Y)$ resembles the MRI of a healthy control. 

Based on existing image-translation methods, one would apply a  binary training strategy \cite{zhu2017unpaired} to learn $\GG$ and $\FF$; i.e., to fool the classifier such that $\PP(\hat{X})>0$ and $\PP(\hat{Y})<0$. However, the neurological condition linked to a brain disorder may lie in a continuous spectrum as encoded in the predicted logit $p$. For example, the severity of cognitive impairment can be highly heterogeneous in the AD cohort such that some AD patients should have larger logit values and others having logits closer to 0. As such, the above binary objective may overemphasize the severe AD cases in converting them into controls, thereby implicitly reweighing the importance across subjects during training. To avoid such bias, we enforce the simulators to produce a \textit{logit shift} greater than a pre-defined threshold $\delta$ for each subject; i.e., $\PP(\hat{X})-\PP(X)>\delta$ and  $\PP(\hat{Y})-\PP(Y)<-\delta$. This logit shift loss is then formulated as
\begin{equation}
    E_{logit} := \mathbf{E}_{X \sim \mathcal{X}} [\text{max}(\PP(X)-\PP(\hat{X}),-\delta)] + \mathbf{E}_{Y \sim \mathcal{Y}}[ \text{max}(\PP(\hat{Y})-\PP(Y),-\delta)].
\end{equation}
As commonly explored in the literature, we also incorporate the \textit{cycle-consistent} loss to ensure that the simulators can recover the original input from a simulated image. This guarantees that subject-specific information in the image irrelevant to the disease is not perturbed during the simulation, i.e. 
\begin{equation}
    E_{cycle} := \mathbf{E}_{X \sim \mathcal{X}} [||\FF(\GG(X)) - X)||_2] + \mathbf{E}_{Y \sim \mathcal{Y}}[||\GG(\FF(Y)) - Y)||_2].
    \label{eq:cycle}
\end{equation}

\subsection{Coupling Simulators via Conditional Convolution}
\begin{figure}[!t]
\centering
\includegraphics[width=\textwidth,trim=10 170 85 105,clip]{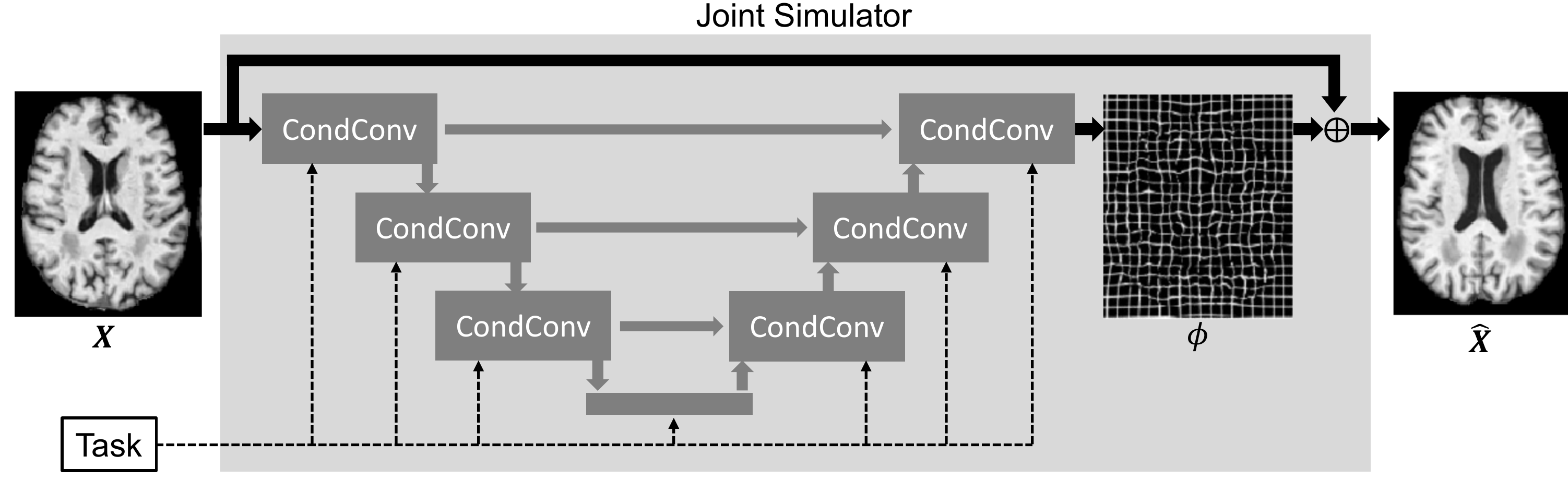}
\caption{Two simulators are coupled into one single model by using Conditional Convolution (CondConv), whose parameters are dependent on the specific simulation task. The output of the joint simulator is a warping field $\phi$ that is applied to the input $X$ to derive the simulated $\hat{X}$.} 
\label{fig:condconv}
\end{figure}

A drawback of traditional cycle-consistent learning is that the two simulators $\GG$ and $\FF$ are designed as independent networks albeit the two simulation tasks are extremely coupled (injecting vs removing disease patterns). In other words, the network parameters between $\GG$ and $\FF$ should be highly dependent, and each convolutional kernel at a certain layer should perform related functions. Here, we propose to combine the two simulators into a coherent model whose behavior can be adjusted based on the specific simulation task. We do so by using \textit{conditional convolution} (CondConv) \cite{yang2019condconv} as the fundamental building blocks of the network shown in Fig. \ref{fig:condconv}. Let $f$ and $f'$ be the input and output features of a convolutional operation with activation $\boldsymbol{\sigma}$. As opposed to the static convolutional kernel, the CondConv kernel $W$ is conditionally parameterized as a mixture of experts
\begin{equation}
    f' := \boldsymbol{\sigma}(\alpha_1 \cdot W_1 \circledast f + ... + \alpha_K \cdot W_K \circledast f), 
\end{equation}
where $W$ is a linear combination of $k$ sub-kernels with weights $\{\alpha_k|k=1,...,K\}$ determined via a \text{routing function} $r_k(\cdot)$. With $t$ being the task label (e.g., 0 for $\GG$ and 1 for $\FF$), we design the following routing function
\begin{equation}
    \alpha_k=r_k(f,t) := \mathtt{sigmoid}([\mathtt{GlobalAvgPool}(f),t] * R_k),
\end{equation}
where $R_k$ are the learnable parameters to be multiplied with the concatenation of the pooled feature and the task label. In doing so, the behavior of the convolution can adapt to subject-specific brain appearance  encoded in $f$ and the specific task.

\subsection{Learning Warping Fields}
In principle, the two simulators can generate any patterns that separate the healthy and diseased cohorts (intensity difference, shape change, etc). In scenarios where the disease is known to impact brain morphometry, we can enforce the simulators to only learn the shape patterns that differentiate cohorts. To do so, we let the output of the simulators be 3D warping fields $\phi_1:=\GG(X)$ and $\phi_2:=\FF(X)$, which are then applied to the input images to derive the warped images $\hat{X} = X \circ \phi_1$, $\hat{Y} = Y \circ \phi_2$. The warping layer is implemented the same as in \cite{voxelmorph}, which uses linear interpolation to compute the intensity of a sub-voxel defined at non-integer locations. As also adopted in \cite{voxelmorph}, a diffusion regularizer is added to the warping field $\phi$ to preserve the smoothness of $\phi$. Let \textbf{V} be the voxels in the image space. The smoothness loss is
\begin{equation}
    E_\phi:=\lambda_{\phi}\sum_{\textbf{v} \in \textbf{V}}|| \nabla \phi(\textbf{v})||^2 \text{ , where } \nabla \phi(\textbf{v})=(\frac{\partial \phi(\textbf{v})}{\partial x},\frac{\partial \phi(\textbf{v})}{\partial y},\frac{\partial \phi(\textbf{v})}{\partial z}).
\end{equation}
The final objective function is  $E:=E_{logit}+E_{cycle}+E_\phi$.

\section{Experiments}
To showcase the concept of our cycle-consistent image simulation, we first evaluated the method on a synthetic dataset by only considering $E_{logit}$ and $E_{cycle}$ during training. We then incorporated $E_\phi$ in the training to show the advantage of warping-field visualization over existing visualization techniques in the context of analyzing Alzheimer's Disease. Lastly, we applied the proposed approach to identify regional atrophy linked to alcohol dependence.

\subsection{Synthetic Experiments} 
\begin{figure}[!t]
\centering
\includegraphics[width=\textwidth,trim=100 60 75 80,clip]{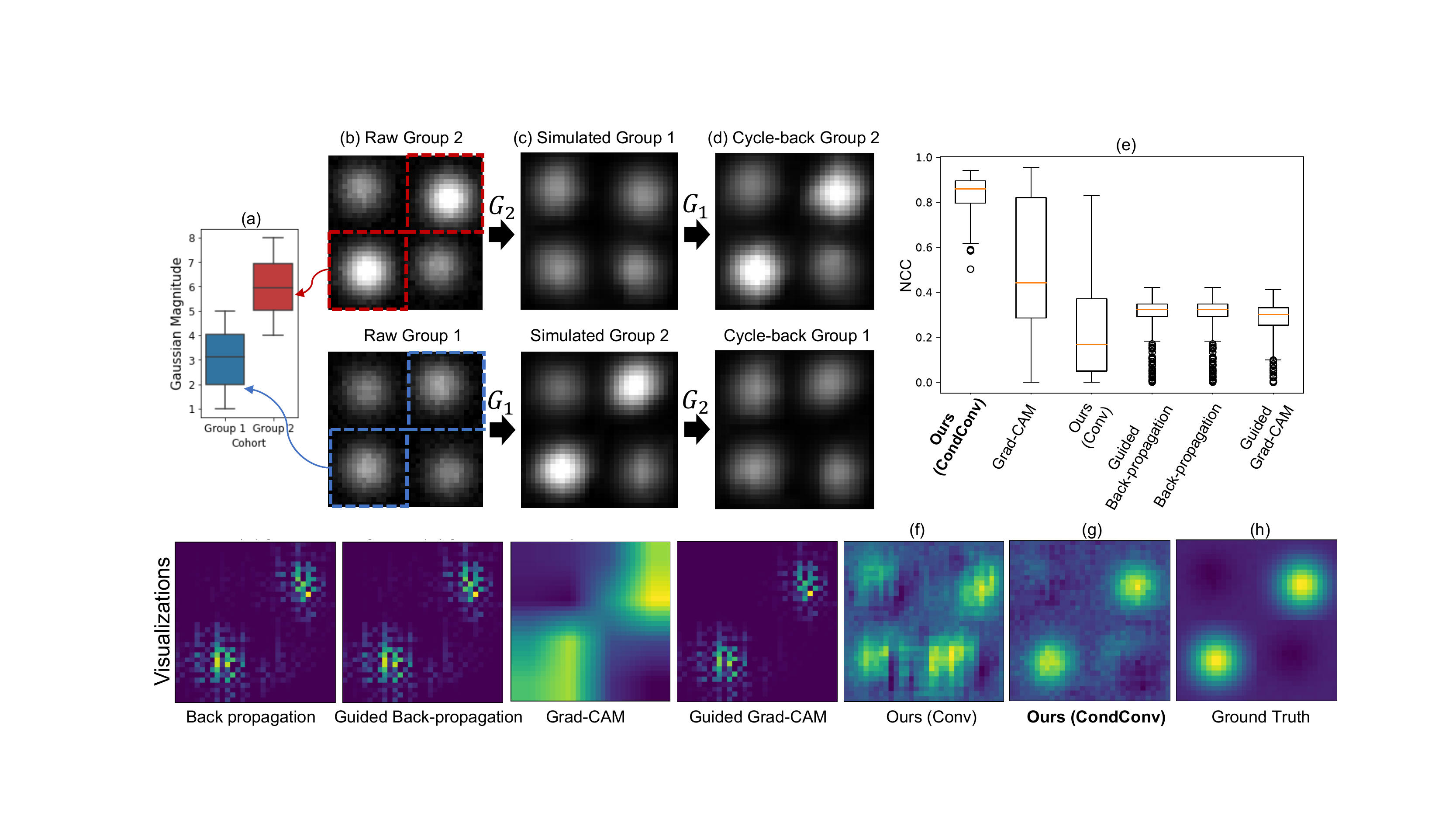}
\caption{(a) The group-separating pattern was the magnitude of two off-diagonal Gaussians. (b,c,d) The learned simulators could reduce and increase the intensity of off-diagonal Gaussians of a given image in a cycle-consistent fashion; (e) Normalized cross-correlation (NCC) between the ground-truth pattern and the pattern derived by different approaches. Bottom: visualizations of the group-separating patterns.  }
\label{fig:synthetic}
\end{figure}
\textbf{Dataset:} We generated a synthetic dataset comprising two groups of data, each containing 512 images of resolution $32\times32$ pixels. Each image was generated by 4 Gaussians, whose locations randomly varied within each of the 4 blocks (Fig. \ref{fig:synthetic}(b)). We assume the magnitude of the two off-diagonal Gaussians defined the differential pattern between the two cohorts. Specifically, the magnitude was sampled from a uniform distribution $\mathcal{U}(1,5)$ for each image from Group 1 and from $\mathcal{U}(4,8)$ (with stronger intensities) for Group 2 (Fig. \ref{fig:synthetic}(a)). On the other hand, the magnitude of the two diagonal Gaussians was sampled from $\mathcal{U}(1,6)$ and regarded as subject-specific information impartial to group assignment. Gaussian noise was added to the images with standard deviation $0.002$. 

\noindent\textbf{Classification:} We first trained a classifier to distinguish the two groups on 80\% of the data. The classifier network ($\PP$) comprised of 3 stacks of 2D convolution (feature dimension  $=\{2,4,8\}$), ReLU, and max-pooling layers. The resulting 128 features were fed to a multi-layer perceptron with one hidden layer of dimension 16 and ReLU activation. Training the classifier resulted in a classification accuracy of 87.5\% on the remaining 20\% testing images.  

\noindent\textbf{Visualization:} To visualize the group-separating pattern learned by the classifier, we designed the simulator as a U-net structure with skip connections (Fig. \ref{fig:condconv}). The encoder was 4 stacks of 2D CondConv (feature dimension  $=\{1,2,4,8\}$), BatchNorm, LeakyReLu, and max-pooling layers. Each CondConv operation used 3 experts ($K=3$) as adopted in the original implementation of \cite{yang2019condconv}. The resulting 64 features were fed into a fully connected layer of dimension 64 and ReLU activation. The decoder had an inverse structure of the encoder by replacing the pooling layers with up-sampling layers. The warping field was not used in this experiment, so the networks directly generated simulated images. The logit shift threshold was set to $\delta=5$. For each test image, we then computed the intensity difference between the raw and simulated images ($X-\hat{X}$). 

\noindent\textbf{Baseline:} We also generated visualizations through 4 baseline approaches: back-propagation (BP), guided BP \cite{springenberg2014striving}, Grad-CAM, and guided Grad-CAM \cite{selvaraju2017grad}. To show the importance of using conditional convolution for our model, we also generated the pattern using our model trained with two separate encoders using conventional convolution. As the results of different approaches had different scales, each estimated pattern was compared with the ground-truth using normalized cross-correlation (NCC), where the ground-truth was defined as the magnitude difference associated with the two off-diagonal Gaussians (Fig. \ref{fig:synthetic}(h)).

\noindent\textbf{Results:} Fig. \ref{fig:synthetic} shows two examples of the learned simulation. For a training image from Group 2 (Fig. \ref{fig:synthetic}(b)), the simulator reduced the intensity of off-diagonal Gaussians, indicating that the model successfully captured the group-separating patterns (Fig. \ref{fig:synthetic}(c)). Meanwhile, the model preserved subject-specific information including the location and magnitude of the two diagonal Gaussians. Through cycle-consistent simulation, the model also accurately recovered the input image (Fig. \ref{fig:synthetic}(d)). In line with the visual comparison, the pattern generated by our model (Fig. \ref{fig:synthetic}(g)) only focused on the off-diagonal Gaussians and closely resembled the ground-truth (Fig. \ref{fig:synthetic}(h)). Note, when replacing the CondConv with conventional convolution, the pattern became less robust (Fig. \ref{fig:synthetic}(f)). On the other hand, the visualizations derived by BP, guided BP and guided Grad-CAM were noisy as the saliency values frequently switched signs. This behavior was inconsistent with our data construction, where the magnitude change of the Gaussians had the same sign at each voxel. The pattern associated with Grad-CAM was too smooth to accurately locate the object of interest. Lastly, this qualitative analysis was supported by the NCC metric (Fig. \ref{fig:synthetic}(e)) indicating our model with CondConv was the most accurate approach for defining the pattern.

\subsection{Visualizing the Effect of Alzheimer's Disease}

\begin{figure}[!t]
\centering
\includegraphics[width=\textwidth,trim=30 140 30 120,clip]{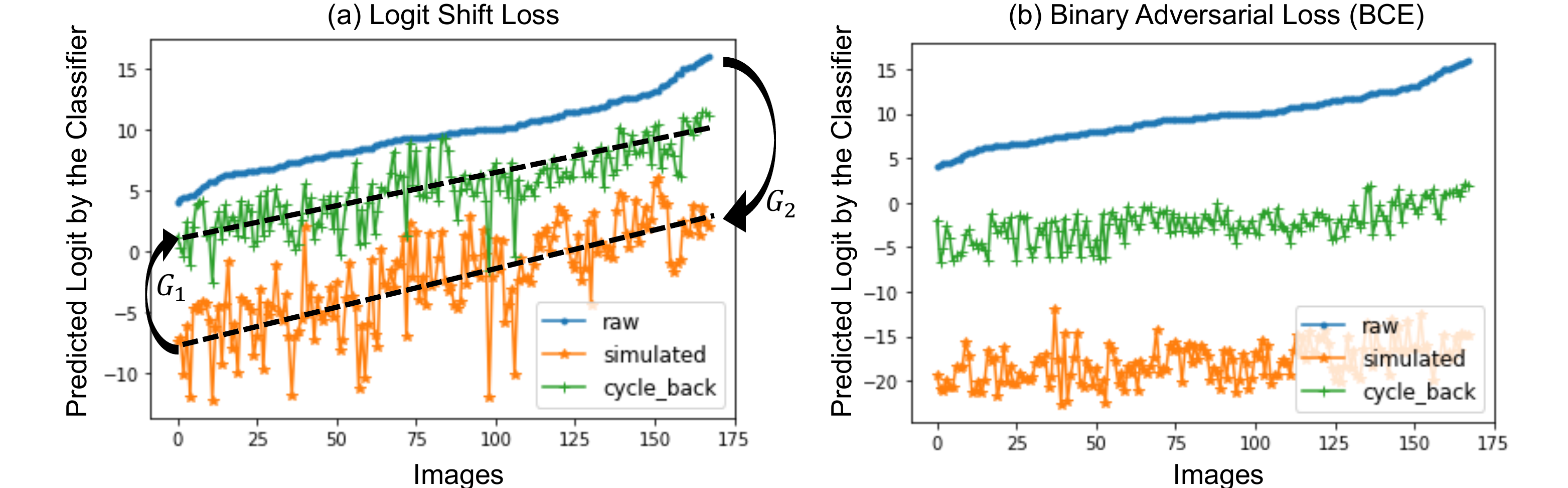}

\caption{Predicted logic values by the classifier of all raw AD images (blue) in the test set, simulated images after removing disease patterns (orange), and cycle-back simulations (green). Images are re-ordered based on the raw logic values. The simulator is learned based on (a) the proposed logit loss or (b) the binary cross-entropy loss .} 
\label{fig:logit}
\end{figure}

\noindent\textbf{Dataset:} We evaluated the proposed model on 1344 T1-weighted MRIs from the Alzheimer’s Disease Neuroimaging Initiative (ADNI1). The dataset consisted of images from 229 Normal Control (NC) subjects (age: 76 $\pm$ 5.0 years) and 185 subjects with Alzheimer’s Disease (75.3 $\pm$ 7.6 years). Each subject had 1 to 8 longitudinal scans within a 4 year study period and only contained MRIs that were successfully preprocessed. The preprocessing consisted of denoising, bias field correction, skull stripping, affine registration to a template, re-scaling to a 64 $\times$ 64 $\times$ 64 volume, and transforming image intensities within the brain mask to z-scores. This dataset was randomly split into 80\% training and 20\% testing on the subject level. 

\noindent\textbf{Implementation:}\footnote{Source code can be found at \url{https://github.com/ZucksLiu/DeepInterpret}.} We first trained a classifier $\PP$ containing 4 stacks of $3\times 3 \times 3$ convolutional layers (feature dimension $\{16,32,64,16\}$), ReLU, and max-pooling layers. The resulting 512 features were fed into a multi-layer perceptron with one hidden layer of dimension 64 and ReLU activation. Based on this architecture, the classifier achieved 88\% NC/AD classification accuracy (balanced accuracy) on the testing set. Note, as the goal of our work was to visualize a trained classifier as opposed to optimizing the classification accuracy on a particular dataset, we did not consider the dependency of longitudinal scans for simplicity. To interpret this trained classifier, we adopted a similar simulator architecture as in the synthetic experiment while using 5 convolutional stacks with 3D CondConv (feature dimension $=\{16,32,64,16,16\}$), a fully connected layer of dimension 512, and a 3-channel output (warping field $\phi$). We set $\lambda_\phi=0.02$ and the logit shift threshold $\delta=12.5$. The simulators were trained on all the NC and AD subjects by an Adam optimizer for 45 epochs with a learning rate of 1e-4.

\noindent\textbf{Results:} We first show the impact of the logit shift loss on the cycle-consistency of the simulation. Fig. \ref{fig:logit}(a) displays the logit values of all raw testing images predicted by the classifier $\PP$ (blue curve). After removing and injecting disease patterns through the cycle-consistent simulators, the logit values consistently decreased and increased while preserving their relative positions. However, if we replaced the logit shift loss by the binary cross-entropy loss \cite{zhu2017unpaired} (BCE, Fig. \ref{fig:logit}(b)), the logit values of the simulated and cycle-back images became all uniform. This was undesirable as the goal of the simulator was to uncover the pattern that correlated with the severity of the brain disorder, which was encoded by the magnitude of the logit values. Using the BCE loss simply  `fooled' the classifier but lost this important information.

\begin{figure}[!t]
\centering
\includegraphics[width=\textwidth,trim=62 95 78 67,clip]{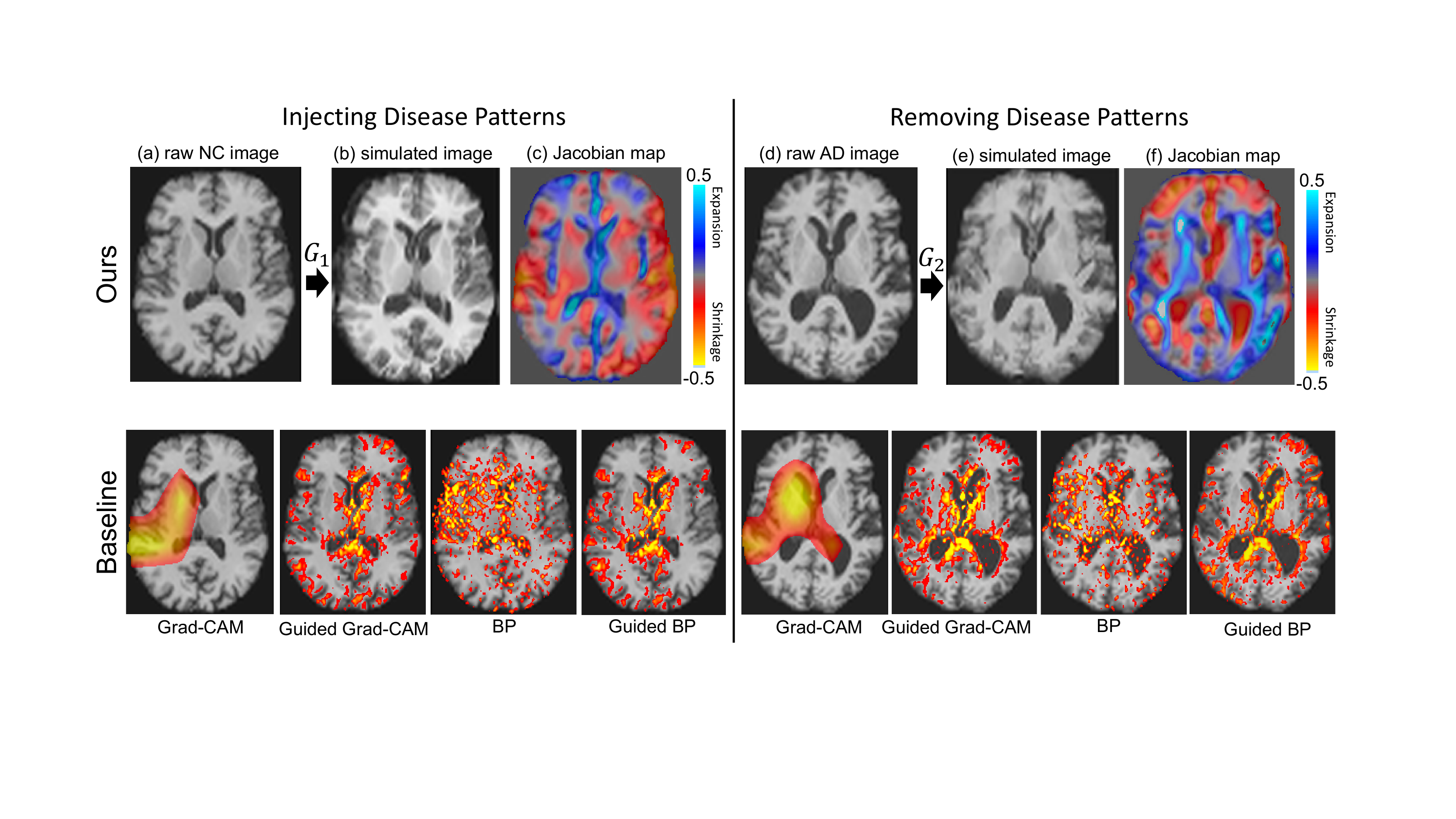}
\caption{Example visualization of our proposed approach and the baselines. Color scales were omitted for baseline approaches as they are arbitrary across models and subjects. All the computation was performed in the down-sampled space and resized to the original resolution for easy visualization.} 
\label{fig:adni1}
\end{figure}

Fig. \ref{fig:adni1}(b) shows a simulated image after injecting the AD pattern into the raw image of an NC subject. By directly comparing the two grayscale images, we observe enlargement of the ventricles and cerebrospinal fluid (CSF) and atrophy of brain tissue. This pattern comported with the effects of AD reported in prior literature \cite{ad}. Moreover, the morphological change captured by the simulator can be quantitatively measured by the log of Jacobian determinant of the warping field $\phi$ (Fig. \ref{fig:adni1}(c)). We used this Jacobian map as the visualization produced by our method, which was then compared with the visualization of the same subject produced by the baseline approaches. In line with the synthetic experiment, the Grad-CAM saliency map was smooth and did not locate meaningful regions with respect to the AD effect. Other saliency maps by BP, guided BP, and guided Grad-CAM were noisy and contained frequent sign changes in the saliency values.  As a second example, we also visualize the simulated image (Fig. \ref{fig:adni1}(e)) after removing the disease pattern from an AD subject and the corresponding Jacobian map (Fig. \ref{fig:adni1}(f)). The patterns were similar to Fig. \ref{fig:adni1}(c) except for the change of direction (regions with shrinkage now showed expansion), indicating the cycle-consistent nature of the two coupled simulators. 

\begin{figure}[!t]
\centering
\includegraphics[width=\textwidth,trim=80 60 110 0,clip]{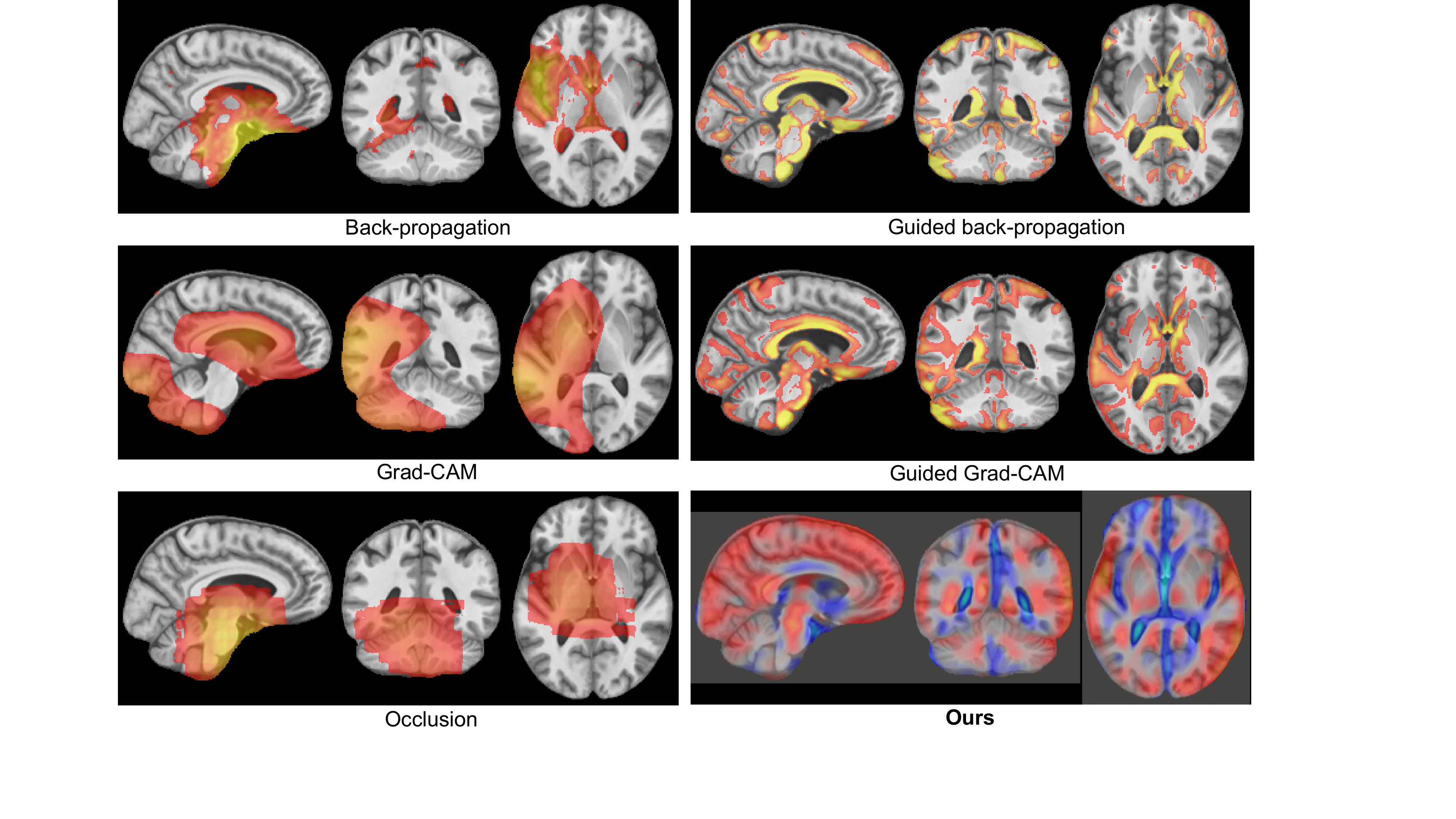}

\caption{Group-level visualization for the ADNI dataset. Color scale is defined according to the subject-level visualization in Fig. 5.} 
\label{fig:adni_all}
\end{figure}

Beyond the subject-level visualization, we also produced the Jacobian visualization on the group level by non-rigidly registering the structural maps of all NC subjects to a template and computing the average Jacobian map in the template space (Fig. \ref{fig:adni_all}). This procedure was also used to produce the group-level visualization of baseline approaches. We also generated a group-level visualization based on the occlusion method \cite{Zeiler2013} (this method cannot be applied to generate subject-level visualization), which first used a sliding window of $8 \times 8 \times 8$ with stride 4 to mask the test images and then re-computed the testing accuracy as the saliency score associated with the center voxel of the sliding window.

After inspecting the subject-level and group-level visualizations, we now summarize 4 advantages of the Jacobian-based visualization: (1) Our Jacobian maps accurately located the disease pattern while avoided generating overly smooth or noisy visualizations; (2) The Jacobian values had real morphological meanings (tissue volume change) while the baseline visualizations only informed the location of the disease pattern; (3) The Jacobian values were signed (shrinkage or expansion) informing the direction of changes, while the sign of the saliency values by baseline methods was less meaningful in the context of MR analysis; (4) The Jacobian values had a deterministic scale (percentage of volume change) while the saliency values of the baseline approaches are meaningless and highly variant across models and subjects.

\subsection{Visualizing the Effect of Alcohol Dependence}
\begin{figure}[!t]
\centering
\includegraphics[width=\textwidth,trim=10 150 30 150,clip]{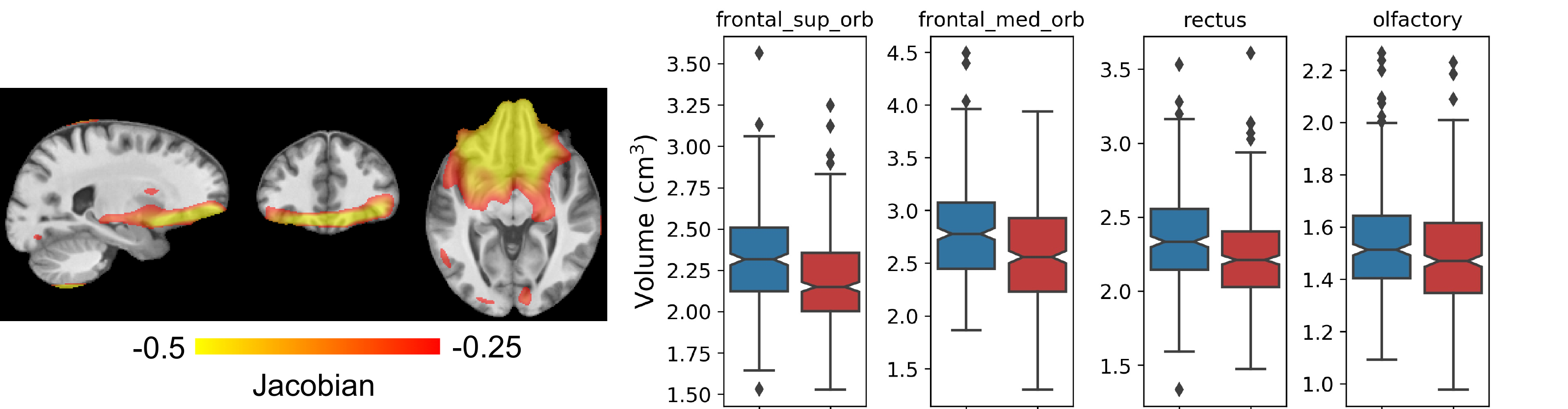}

\caption{Left: Jacobian visualization of the effect of alcohol dependence; Right: gray matter volume of 4 brain regions from the orbitofrontal cortex measured for all NC and alcohol-dependent subjects. The volumes scores were corrected for age and sex.} 
\label{fig:AUD}
\end{figure}

The dataset comprised 1225 T1-weighted MRIs of 274 NC subjects (age: $47.3 \pm 17.6$) and 329 participants with alcohol dependence (age: $49.3 \pm 10.5$) according to DSM-IV criteria \cite{Sullivan2018}. 74 of the alcohol dependent group were also human immunodeficiency virus (HIV) positive. All experimental settings replicated the ADNI experiment. Based on an 80\%-20\% training and testing split on the subject level, the classifier resulted in a 76.2\% accuracy for classifying alcohol-dependent subjects. After training the simulators on all images, we computed the group-level Jacobian visualization for all NC subjects in the testing set. Fig. \ref{fig:AUD} indicates that regions with the most severe atrophy were located in the orbitofrontal cortex. This converges with recent studies that frequently suggested the disruption in the structural and functional properties of the orbitofrontal cortex associated with alcohol dependence \cite{Moorman2018}.

To confirm this finding, we tested the group difference in the volumetric measures of 4 regions of interest: the superior and medical fronto-orbital gyri, the rectus, and the olfactory gyrus. Only the baseline MR of each subject was used in this analysis. The volumetric measures were extracted by segmenting the brain tissue into gray matter, white matter, and CSF via Atropos and parcellating the regions by the SRI24 atlas.  With age and gender being the covariates, a general linear model tested the group difference between NC and alcohol-dependent subjects in the gray matter volume of the 4 regions. All tests resulted in significant group differences based on two-tailed t-statistics ($p<0.001$, see boxplots in Fig. \ref{fig:AUD}), which confirmed the validity of our visualization result. These results indicate that our data-driven visualization can be readily combined with \textit{a priori} regional analysis (e.g., Fig. \ref{fig:AUD}), which allows neuroscientists to cross-validate findings from data-driven and hypothesis-driven analyses.

\section{Conclusion}
In this work, we have proposed a novel interpretation/visualization technique based on the image-to-image translation scheme. By learning simulators that could inject and remove disease patterns from a given image, our approach permitted human-understandable and robust visualization on both the subject level and group level.
While the experiment focused on identifying morphological changes associated with a disease, the proposed framework has the potential to study generic disease effects, e.g., intensity changes induced by lesions. 
Moreover, our method also has great generalizability as it is independent of the classifier's architecture.  In summary, our work marks an important step towards the application of deep learning in neuroimaging studies.

\textbf{Acknowledgement.} This work was supported by NIH Grants MH113406, AA005965, AA010723, and AA017347, and by Stanford HAI AWS Cloud Credit. 
%
%
%
\bibliographystyle{splncs}
\bibliography{ipmi}

\end{document}